\documentclass{xjkas}

\def\year{2016} 
\def\month{September} 
\def\volume{00} 
\def\beginpage{1} 
\def\endpage{6} 
\def\received{2016 July 19} 
\def\accepted{2016 September 26} 
\date{Received \received; accepted \accepted}

\usepackage{balance}

\usepackage[breaklinks,colorlinks,citecolor=blue,linkcolor=blue,menucolor=blue,urlcolor=blue]{hyperref}

\def\msol{M_{\odot}}

\def\kms{{\rm km\,s}^{-1}}
\def\ke{\langle k_e \rangle}
\def\ka{\langle k_a \rangle}
\def\atlas{ATLAS$^{\rm 3D}$}

\title{
The Virial Relation and Intrinsic Shape of Early-Type Galaxies
}

\author[]{Sascha Trippe}

\affil[]{Department of Physics and Astronomy, Seoul National University, Seoul 08826, Korea; \email{trippe@astro.snu.ac.kr}}

\def\corrauthor{
S. Trippe
}

\def\runningauthor{
Trippe
}

\def\runningtitle{
The Virial Relation and Intrinsic Shape of ETGs
}

\def\keywords{
galaxies: elliptical and lenticular, cD --- galaxies: kinematics and dynamics --- galaxies: fundamental parameters --- galaxies: structure
}

\def\abstracttext{
Early-type galaxies (ETGs) are supposed to follow the virial relation $M = k_e \sigma_*^2 R_e / G$, with $M$ being the mass, $\sigma_*$ being the stellar velocity dispersion, $R_e$ being the effective radius, $G$ being Newton's constant, and $k_e$ being the virial factor, a geometry factor of order unity. Applying this relation to (a) the \atlas\ sample of \citet{cappellari2013a} and (b) the sample of \cite{saglia2016} gives ensemble-averaged factors $\ke=5.15\pm0.09$ and $\ke=4.01\pm0.18$, respectively, with the difference arising from different definitions of effective velocity dispersions. The two datasets reveal a statistically significant tilt of the empirical relation relative to the theoretical virial relation such that $M\propto(\sigma_*^2R_e)^{0.92}$. This tilt disappears when replacing $R_e$ with the semi-major axis of the projected half-light ellipse, $a$. All best-fit scaling relations show zero intrinsic scatter, implying that the mass plane of ETGs is fully determined by the virial relation. Whenever a comparison is possible, my results are consistent with, and confirm, the results by \citet{cappellari2013a}. The difference between the relations using either $a$ or $R_e$ arises from a known lack of highly elliptical high-mass galaxies; this leads to a scaling $(1-\epsilon) \propto M^{0.12}$, with $\epsilon$ being the ellipticity and $R_e = a\sqrt{1-\epsilon}$. Accordingly, $a$, not $R_e$, is the correct proxy for the scale radius of ETGs. By geometry, this implies that early-type galaxies are axisymmetric and oblate in general, in agreement with published results from modeling based on kinematics and light distributions.
}

\begin{document}
\jkashead 


\section{Introduction \label{sec:intro}}

Early-type galaxies (ETGs) are known to follow characteristic scaling relations between several structural parameters including, most prominently, the line-of-sight velocity dispersion $\sigma_*$ of their stars. The first of these relations to be discovered was the Faber--Jackson relation $L\propto\sigma_*^4$ between velocity dispersion and galactic luminosity $L$ \citep{faber1976}. Subsequent work led to the discovery of the ``classic'' fundamental plane relation $L\propto\sigma_*^{\alpha}I^{\beta}$, or equivalently, $R_e\propto\sigma_*^{\gamma}I^{\delta}$ which include the (two-dimensional, projected) effective (half-light) radius $R_e$ and the average surface brightness $I$ within $R_e$ \citep{dressler1987, djorgovski1987}; recent studies \citep{cappellari2013a} find $\gamma\approx1$ and $\delta\approx-0.8$. From the virial relation $M\propto\sigma_*^2R_e$, with $M$ being the galaxy mass, one expects $\gamma=2$ and $\delta=-1$; the tilt of the fundamental plane, i.e. the discrepancy between expected and observed values, can be ascribed to a scaling of mass-to-light ratio with velocity dispersion (\citealt{cappellari2006}; but see also \citealt{cardone2011}). The Faber--Jackson relation can be understood as a projection of the fundamental plane onto the $L$--$\sigma_*$ plane (but see also \citealt{sanders2010}).

Whereas scaling relations that involve $L$ are convenient because the luminosity is an observable, the galactic dynamics is controlled by the galaxy mass $M$ for which $L$ is a proxy. For pressure-supported stellar systems, mass and velocity dispersion are connected via the virial relation
\begin{equation}
\label{eq:virial}
M = k_e\,\frac{\sigma_*^2\,R_e}{G}
\end{equation}
where $G$ is Newton's constant and $k_e$ is a geometry factor of order unity \citep[e.g.,][]{binney2008}. Accordingly, a ``more fundamental plane'' \citep{bolton2007} is given by the ``mass plane'' relation $M\propto\sigma_*^{\kappa}R_e^{\lambda}$; from the virial theorem, one expects $\kappa=2$ and $\lambda=1$ (see also \citealt{cappellari2016} for a recent review).

\begin{figure*}[t!]
\centering
\includegraphics[angle=-90, width=84mm]{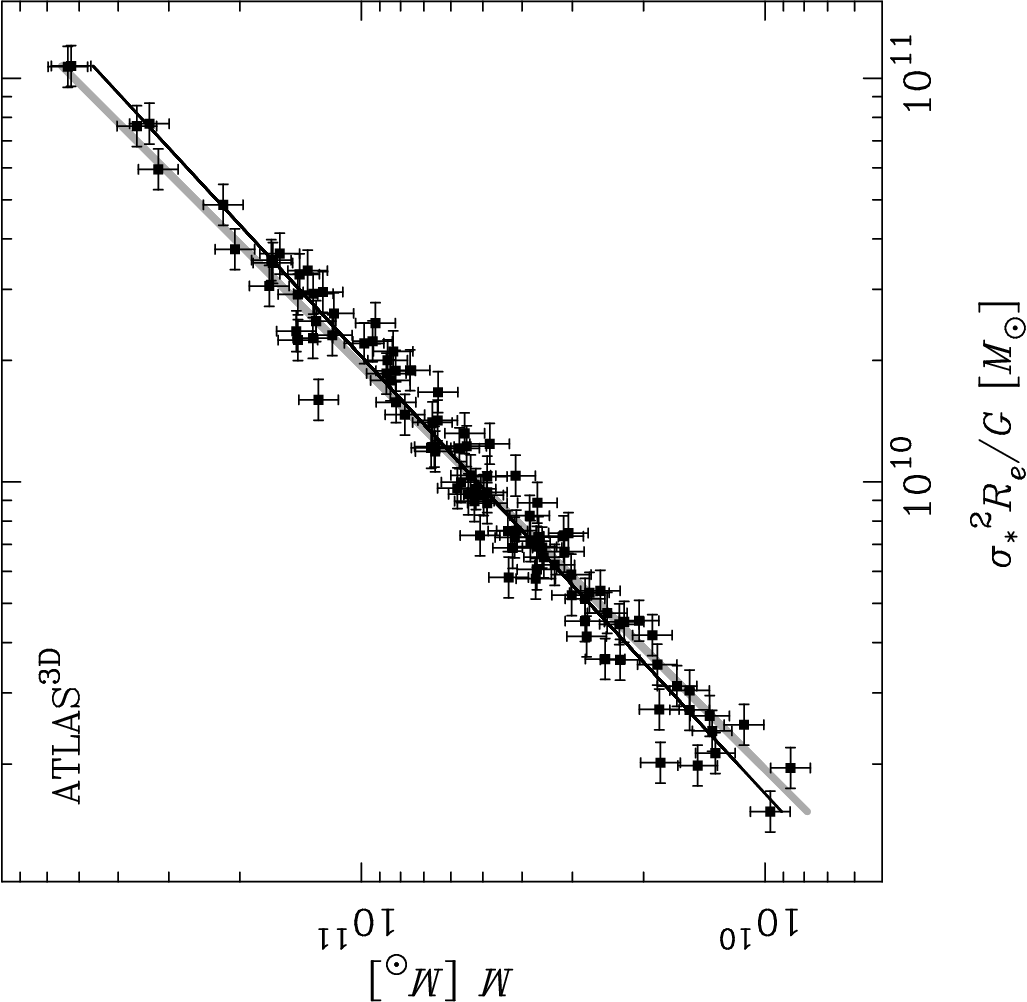} 
\hspace{4mm} 
\includegraphics[angle=-90, width=84mm]{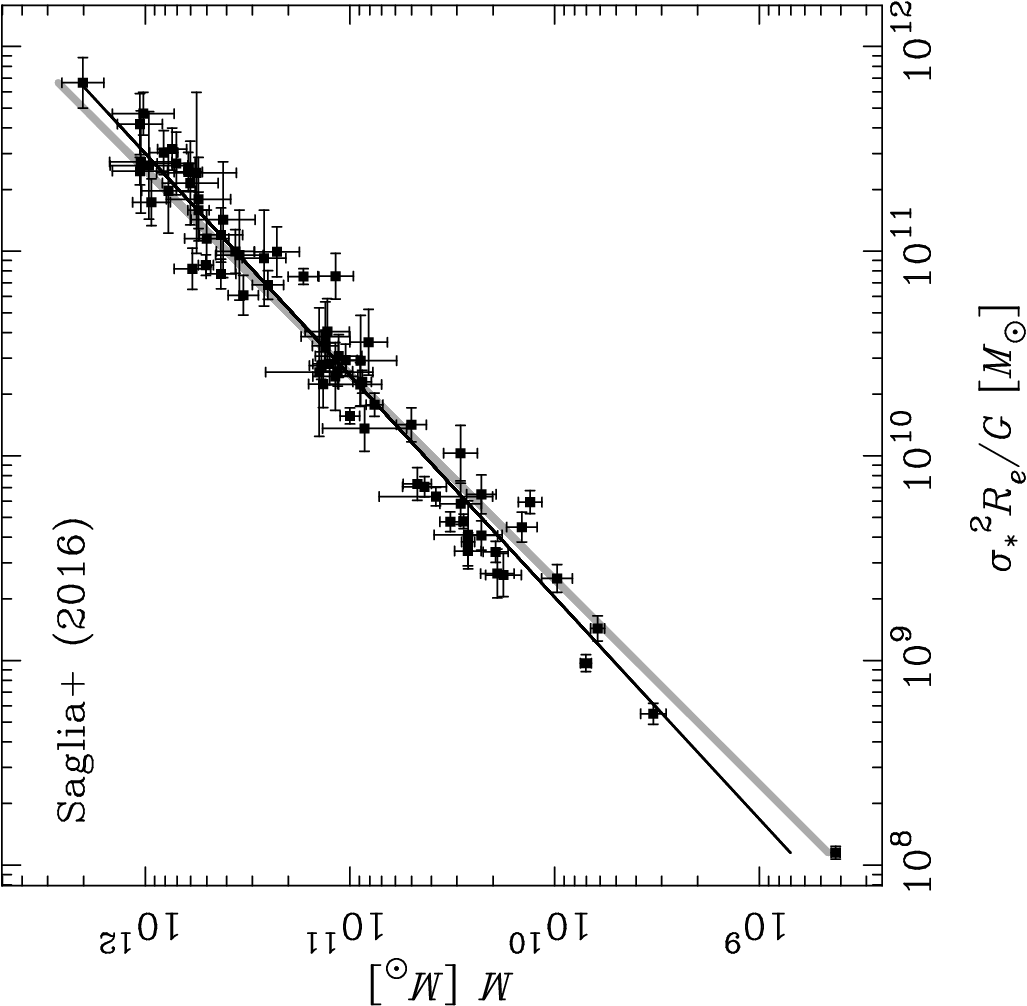}
\caption{Galaxy mass $M$ as function of virial term $\sigma_*^2R_e/G$, both in units of solar mass. Please note the somewhat different axis scales. \emph{Left:} for the \atlas\ sample. The grey line corresponds to a linear relation (Equation \ref{eq:virial}) with ensemble-averaged virial factor $\ke = 5.15$. The black line marks the best-fit generalized virial relation (Equation~\ref{eq:massplane}) with $x = 0.924 \pm 0.016$ \emph{Right:} for the data of \citet{saglia2016}. The grey line indicates a linear relation with $\ke = 4.01$. The black line marks the best-fit generalized virial relation with $x = 0.923 \pm 0.018$. \label{fig:virialRe}}
\end{figure*}

Testing the validity and accuracy of Equation~(\ref{eq:virial}) is important as virial mass estimators of this type are widely applied to pressure-supported stellar systems. Fundamental plane studies usually derive masses $M$ from photometry and equate them with dynamical masses, presuming an equality of the two. A deviation from the theoretical relation would imply the presence of additional ``hidden'' parameters or dependencies between parameters. Else than for relations between luminosity and other parameters, a tilt in the mass plane would be connected immediately to the dynamics or structure of galaxies. To date, the virial relation (Equation~\ref{eq:virial}) is commonly assumed to be valid exactly (cf., e.g., \citealt{cappellari2013a}). This is, however, not undisputed. Based on an analysis of about 50\,000 SDSS galaxies, \citet{hyde2009} concluded that $M\propto(\sigma_*^2R_e)^{0.83\pm0.01}$. More recent observations, accompanied by more sophisticated dynamical modeling, of early-type galaxies in three nearby galaxy clusters find $\kappa\approx1.7$ \citep{scott2015}. This raises the question to what extend Equation~(\ref{eq:virial}) is appropriate for describing the dynamics of ETGs, and which alternative formulations might be necessary.

\section{Data \label{sec:data}}

This work is primarily based on the \atlas\ database of \citet{cappellari2011, cappellari2013a}. In addition, I use the dataset of \citet{saglia2016} for an independent check. The two samples cannot be combined directly because they employ different conventions for calculating effective stellar velocity dispersions.

\subsection{The ATLAS$^{\rm\bf 3D}$ Sample \label{sec:atlasd}}

The \atlas\ project \citep{cappellari2011, cappellari2013a} provides\footnote{\url{http://www-astro.physics.ox.ac.uk/atlas3d}} data for a volume-limited sample of 260 nearby (located within $\lesssim$42 Mpc) early-type galaxies. For each galaxy, the surface brightness distribution is modeled with a Multi-Gaussian Expansion (MGE) algorithm. The results are fed into an Jeans Anisotropic MGE (JAM) algorithm which computes predictions for the line-of-sight velocity dispersion distribution in the sky plane. These values are compared to observed velocity dispersion distributions obtained from optical integral-field spectroscopy. The best-fit JAM models provide the effective radius $R_e$ and masses $M$. For each galaxy, an effective velocity dispersion $\sigma_*$ is measured from a combined spectrum co-added over an ellipse of area $\pi R_e^2$.

As the JAM results vary in quality, some quality-based selection of data is needed. Following the suggestion of \citet{cappellari2013a}, I select galaxies for which there is at least a ``good'' (quality flag $\geq$2) agreement between predicted and observed velocity dispersion distributions. This results in a final dataset comprising 101 galaxies. The selected galaxies have (JAM) masses $M$ between $\approx$$9\times10^{9}\,\msol$ and $\approx$$5\times10^{11}\,\msol$, effective velocity dispersions $\sigma_*$ between $\approx$70~$\kms$ and $\approx$280~$\kms$, and effective radii $R_e$ ranging from $\approx$0.5~kpc to $\approx$7~kpc. Formal uncertainties are 10\% (0.041 dex) for effective radii, 5\% (0.021 dex) for effective velocity dispersions, and 12\% (0.049 dex) for galaxy (JAM) masses.

\subsection{The Sample of \citet{saglia2016} \label{sec:saglia}}

The dataset by \citet{saglia2016} provides data for 72 local (located within $\lesssim$150 Mpc) elliptical galaxies and classical bulges. Classical bulges can be regarded as elliptical galaxies that formed new disks around them; they follow the same parameter correlations as ``free'' ellipticals do \citep{kormendy2012}. Accordingly, I will treat both types of objects jointly from now on. The sample of \citet{saglia2016} was selected for studies of black hole -- host galaxy relations and combines (re-calibrated where necessary) literature results with new integral-field spectroscopic observations.

For each galaxy, mass and scale radius are derived from photometry. Each target is decomposed into its elliptical bulge and other components like disks, rings, or bars (if any). Bulge masses $M$ are calculated from their luminosities using mass-to-light ratios derived from dynamical modeling. The three-dimensional spherical half-mass radii $r_h$ are used as scale radii. The effective stellar velocity dispersion $\sigma_*$ is derived from a brightness-weighted sum of the squares of velocity dispersion and rotation speed over radii from 0 to $R_e$. The radii $r_h$ and $R_e$ are related like $R_e=(0.74\pm0.01)\,r_h$ (\citealt{saglia2016} for their sample; see also \citealt{hernquist1990, wolf2010} for general derivations).

Sample galaxies were selected with emphasis on covering a wide range in $\sigma_*$, from $\approx$70~$\kms$ to $\approx$390~$\kms$. Half-mass radii $r_h$ range from $\approx$0.1~kpc to $\approx$32~kpc, bulge masses $M$ are located between $\approx$$4\times10^8\,\msol$ and $\approx$$2\times10^{12}\,\msol$. Median formal uncertainties are 21\% (0.083 dex) for bulge masses, 5\% (0.021 dex) for effective velocity dispersions, and 25\% (0.096 dex) for half-light radii.

\section{Analysis and Results \label{sec:analysis}}

\begin{figure}[t!]
\centering
\includegraphics[angle=-90, width=84mm]{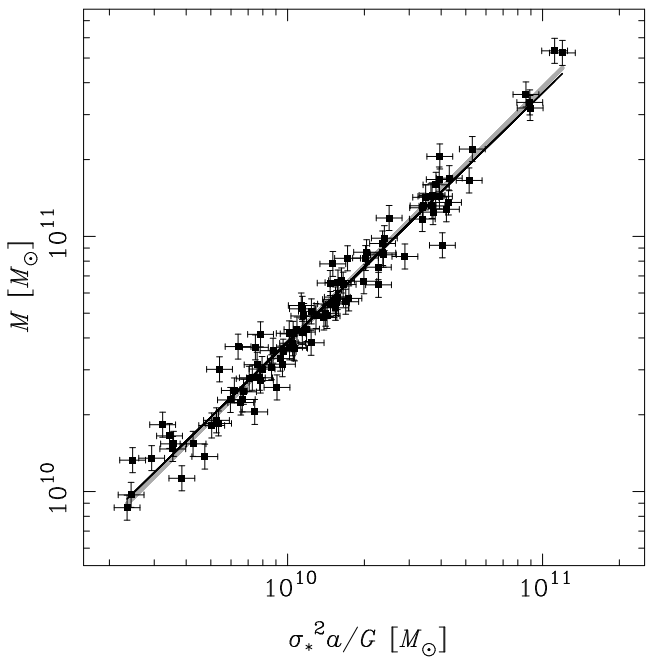} 
\caption{Galaxy mass $M$ as function of virial term $\sigma_*^2a/G$ for the \atlas\ sample, with semi-major axis $a$. The grey line indicates a linear relation (Equation~\ref{eq:virial-a}) with $\ka = 3.82$. The black line marks the best-fit generalized virial relation (Equation~\ref{eq:massplane-a}) with $x' = 0.976 \pm 0.018$. \label{fig:virial-a}}
\end{figure}

\subsection{The Virial Relation \label{sec:virial}}

\subsubsection{Effective Radius as Scale Radius \label{sec:virialRe}}

Masses, velocity dispersions, and radii are (supposed to be) connected via the virial relation expressed by Equation~(\ref{eq:virial}). Figure~\ref{fig:virialRe} shows mass $M$ as function of virial term $\sigma_*^2R_e/G$ for the two samples. Assuming a linear relation gives the best agreement for an ensemble-averaged $\ke = 5.15\pm0.09$ and $\ke = 4.01\pm0.18$ (standard errors of means) for the \atlas\ and \citet{saglia2016} samples, respectively, in full agreement with \citet{cappellari2013a} (for \atlas). Assuming that different conventions for calculating $\sigma_*$ explain the difference entirely, the velocity dispersions of \citet{saglia2016} are systematically higher than the ones of \atlas\ by 13\%.

Taking a closer look however, the data deviate systematically from a naive linear relation. For a quantitative analysis I use the generalized virial relation
\begin{equation}
\label{eq:massplane}
\log\left(\frac{M}{10^{11}\msol}\right) = x\log\left(\frac{\sigma_*^2R_e/G}{10^{10.5}\msol}\right) + y
\end{equation}
where $x$ and $y$ are free parameters; mass and virial term are normalized by their approximate medians in order to minimize the covariance of the fit parameters. Logarithms are decadic. Equation~(\ref{eq:massplane}) describes a ``restricted mass plane'' because $\sigma_*$ and $R$ are coupled like $M\propto\sigma_*^{2x}R^x$ instead of a more general relation $M\propto\sigma_*^{\kappa}R^{\lambda}$ with independent $\kappa$ and $\lambda$. By construction, the restricted mass plane probes the evolution of the ratio of observed and dynamically expected masses.

I fit Equation~(\ref{eq:massplane}) to the data via a standard weighted linear least-squares regression. Error bars are rescaled iteratively such that min$(\chi^2/{\rm d.o.f.})=1$. The best-fit slopes are $x = 0.924\pm0.016$ and $x = 0.923\pm0.018$ (with formal $1\sigma$ errors) for the \atlas\ and \citet{saglia2016} data, respectively. Both values are in good agreement with each other and both are significantly -- by $4.8\sigma$ and $4.3\sigma$, respectively -- smaller than unity: the empirical relation is flatter than the theoretical one. The intrinsic scatter (i.e., the difference in squares of rms residual and bivariate rms measurement error) about the best-fit lines is consistent with zero in both cases.

\subsubsection{Semi-Major Axis Length as Scale Radius \label{sec:virial-a}}

\begin{figure}
\includegraphics[angle=-90, width=84mm]{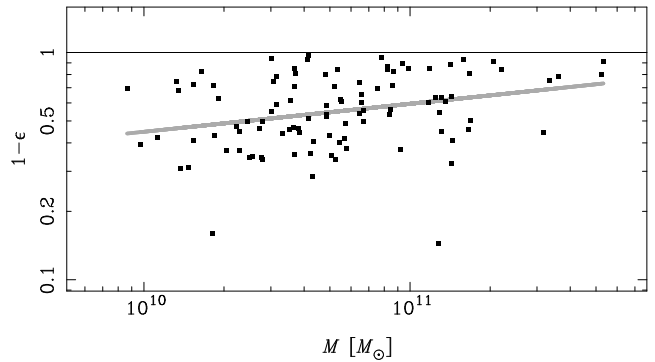}
\caption{Evolution of ``roundness'' $1 - \epsilon$ as function of galaxy mass $M$. The grey line marks the best-fit powerlaw relation, with a slope of $0.123 \pm 0.041$. \label{fig:ell-mass}}
\end{figure}

The effective radius is given by $R_e = \sqrt{a b}$, with $a$ and $b$ being the semi-major and semi-minor axis of the \emph{projected} ellipse that encloses half of the galaxy light, respectively. As argued by, e.g., \citet{hopkins2010}, the semi-major axis $a$ is a more robust proxy for the physical scale radius of a galaxy than $R_e$. Replacing $R_e$ with $a$ results in a modified virial relation
\begin{equation}
\label{eq:virial-a}
M = k_a\,\frac{\sigma_*^2\,a}{G}
\end{equation}
which replaces Equation~(\ref{eq:virial}) and an updated ``restricted mass plane'' relation
\begin{equation}
\label{eq:massplane-a}
\log\left(\frac{M}{10^{11}\msol}\right) = x'\log\left(\frac{\sigma_*^2 a/G}{10^{10.5}\msol}\right) + y'
\end{equation}
which replaces Equation~(\ref{eq:massplane}). By definition, $R_e$ and $a$ are related like $a = R_e / \sqrt{1 - \epsilon}$, with $\epsilon = 1 - b/a$ being the ellipticity. The discussion in the remainder of Section~\ref{sec:analysis} refers to the \atlas\ dataset only because \citet{saglia2016} do not provide ellipticity or semi-major axis length information for their sample galaxies.

Figure~\ref{fig:virial-a} shows galaxy mass $M$ as function of virial term $\sigma_*^2 a / G$, with $a$ computed from $R_e$ and $\epsilon$. Assuming a linear relationship gives an ensemble-averaged virial factor $\ka = 3.82 \pm 0.062$, again in good agreement with \citet{cappellari2013a}. Fitting Equation~(\ref{eq:massplane-a}) to the data (in the same way as done in Section~\ref{sec:virialRe}) results in a slope of $x' = 0.976 \pm 0.018$ -- which agrees with unity within errors. The intrinsic scatter about the best-fit line is consistent with zero.

\subsection{Ellipticity as Function of Mass \label{sec:ell-mass}}

Whereas use of $a$ in the virial relation results in agreement between data and expectation (Section~\ref{sec:virial-a}), use of $R_e$ finds an empirical relation that is significantly flatter than theoretically expected (Section~\ref{sec:virialRe}). As $R_e = a\sqrt{1 - \epsilon}$, the difference between the two empirical relations implies that the ellipticity $\epsilon$ is a function of galaxy mass. Figure~\ref{fig:ell-mass} illustrates the scaling of the ``roundness'' $1-\epsilon$ with $M$ for the \atlas\ sample. For a quantitative test, I fit the relation 
\begin{equation}
\log\left(\frac{1-\epsilon}{0.5}\right) = \xi \log\left(\frac{M}{10^{11}\msol}\right) + \zeta
\label{eq:ell-mass}
\end{equation}
to the data; $\xi$ and $\zeta$ are free parameters. The fit returns a slope of $\xi = 0.123 \pm 0.041$; galaxies of higher mass tend to be less elliptical in average than the ones of lower mass.

\section{Discussion \label{sec:discuss}}

The very existence of a tight mass plane relation is somewhat puzzling. Light distributions, mass-to-light ratios, and thus the masses $M$ of EGS are derived from carefully modeling each system individually. It is not obvious that $M$ should correlate as tightly with the coarse proxies for mass, $\sigma_*$ and either $R_e$ or $a$ (combined in the virial term), as it does, with zero intrinsic scatter (see also the corresponding discussion in \citealt{cappellari2013a}). The global dynamics of ETGs is simpler than one might expect given that they show a wide range of geometries and mass profiles. Likewise, it is noteworthy that a generalized virial relation (Sections \ref{sec:virialRe} and \ref{sec:virial-a}) with slope $x$ or $x'$ describes the dynamics of EGS completely: as there is zero intrinsic scatter about the best-fit lines, adding another free parameter by letting $M$ scale independently with $\sigma_*$ and either $R_e$ or $a$ would not provide additional information (Occam's razor). This is illustrated in Section 4.3 of \citet{cappellari2013a}: their mass plane analysis, using their full sample of galaxies minus a few outliers, returns $M \propto \sigma_*^{1.928} R_e^{0.964}$ which is identical to $M \propto (\sigma_*^2 R_e)^{0.964}$ -- as expected for a fit with too many free parameters.

The analysis in Section~\ref{sec:virialRe} unambiguously shows that, when using $R_e$ as scale radius, the virial relation is tilted, with $x\approx0.92$ being significantly smaller than unity. This is consistent with the original mass plane analysis by \citet{cappellari2013a} who found that their values for $\kappa$ and $\lambda$ were smaller than the expected values by 2.8$\sigma$ and 2.0$\sigma$, respectively. Indeed, when combining the corresponding false alarm probabilities, the analysis by \citet{cappellari2013a} finds a tilt with a significance of $3.7\sigma$ -- but doing this is permissible only when assuming a priori that $\kappa$ and $\lambda$ are correlated (which is at odds with a mass \emph{plane} analysis). My result also qualitatively agrees with the trend observed by \citet{hyde2009}; however, I find a value for $x$ which is significantly larger than the one found from the SDSS sample. Given that I find the same result from two independently drawn and modeled samples of ETGs, I suspect that \citet{hyde2009} underestimated their systematic uncertainties. This can be compared to the results by \cite{scott2015} who found $x = 0.93\pm0.06$ for their sample, which likewise was smaller than unity, however not yet statistically significant. \citet{scott2015} suspected the result $x \neq 1$ to be a feature of JAM modeling. Given however that \citet{saglia2016} use several different types of dynamical modeling to derive EGS masses (cf. their Section~2.2), it seems unlikely that the tilt in the virial relation can be a modeling artifact. 

As shown in Section~\ref{sec:virial-a}, empirical and theoretical virial relation agree (within errors) when adopting the semi-major axis of the half-light ellipse, $a$, as galactic scale radius. Indeed, an improved agreement in this case was already noted in the original mass plane analysis by \citet{cappellari2013a} (cf. also Section 4.2.1 of \citealt{cappellari2016}), although in their analysis the difference between the two formulations (with $R_e$ and $a$, respectively) was not yet statistically significant when assuming independence of $\kappa$ and $\lambda$ from each other. The difference between using $a$ and using $R_e$ in the virial relation arises from a scaling of ellipticity $\epsilon$ with galaxy mass: the higher $M$, the higher the roundness $1-\epsilon$ (Section~\ref{sec:ell-mass}). As noted by \citet{vanderwel2009} and \citet{weijmans2014}, this trend is due to a lack of highly elliptical ($b/a < 0.6$) galaxies at masses $M \gtrsim 10^{11}\msol$ (see also Figure~\ref{fig:ell-mass}); \citet{vanderwel2009} interpreted this observation as evidence for major merging being the dominant mechanism for forming massive galaxies. With $(1-\epsilon) \propto M^{0.12}$ and thus $\sqrt{1-\epsilon} \propto M^{0.06}$, the empirical roundness--mass relation is sufficient to explain the difference between $x\approx0.92$ and unity (within errors). It seems that my analysis is the first to explicitly note the impact of the roundness--mass relation on the virial and mass plane relations of early-type galaxies.

Accepting Equation~(\ref{eq:virial-a}) as the correct virial relation means accepting that $a$ is a proper proxy for the scale radius of early-type galaxies (whereas $R_e$ is not). This was already suggested by \citet{hopkins2010} and later supported by \citet{cappellari2013a}. \citet{hopkins2010} argued that $R_e$ is affected by projection whereas $a$ is not: the same axisymmetric and oblate galaxy viewed under different angles will show different $R_e$ but always the same $a$. (This is actually the reason why \citet{cappellari2013a} argued in favor of using $a$; they did not yet note the effect of ellipticity scaling with ETG mass.) Combining this argument with the fact that Equation~(\ref{eq:virial-a}) fits the available data with no intrinsic scatter implies that \emph{early-type galaxies are intrinsically axisymmetric and oblate in general} -- if they were triaxial or prolate, $a$ would not usually coincide with the longest axis in projection and would not be a measure of galaxy size. For the \atlas\ sample, uncertainties on either $a$ or $R_e$ are given as 10\%, limiting deviations from axisymmetry -- more specifically, the deviation of the ratio of the two longest axes of a triaxial ellipsoid from unity -- to about the same amount. This is in good agreement with the results from modeling the intrinsic shapes of early-type galaxies based on their kinematics and light distributions (with the possible exception of a small sub-population of slowly rotating ETGs; \citealt{weijmans2014}).

\section{Conclusions \label{sec:conclude}}

Using public data for the early-type galaxy samples of \citet{cappellari2011, cappellari2013a} and \citet{saglia2016}, I probe the validity and accuracy of the virial relation given by Equation~(\ref{eq:virial}). The key results are:
\begin{enumerate}
\item Assuming a linear relationship between galaxy mass and virial term, I find ensemble-averaged virial factors of $\ke = 5.15 \pm 0.09$ and $\ke = 4.01 \pm 0.18$ for the \atlas\ and \citet{saglia2016} samples, respectively, in agreement with \citet{cappellari2013a} (for \atlas). The difference between the two samples arguably arises from the \citet{saglia2016} velocity dispersions being systematically higher than the \atlas\ ones by 13\% due to different conventions.
\item For both galaxy samples, the empirical virial relation is significantly (by more than $4\sigma$) tilted, such that $M \propto (\sigma_*^2 R_e)^{0.92}$. For the \atlas\ data, this is consistent with the mass plane analysis provided \citet{cappellari2013a}.
\item Replacing the effective radius $R_e$ with the semi-major axis of the projected half-light ellipse $a$ reconciles empirical and theoretical virial relations, with $M \propto \sigma_*^2 a$ (Equations \ref{eq:virial-a} and \ref{eq:massplane-a}). The ensemble-averaged virial factor is $\ka = 3.82 \pm 0.062$, in good agreement with \citet{cappellari2013a}.
\item All best-fit virial relations show intrinsic scatter consistent with zero. This implies that the mass plane of ETGs is fully determined by the virial relation, i.e., that masses $M$ do not scale independently with $\sigma_*$ and either $R_e$ or $a$ but only with $\sigma_*^2 R_e$ (or $\sigma_*^2 a$).
\item The ``roundness'' $1-\epsilon$, with ellipticity $\epsilon$, mildly scales with galaxy mass such that $(1 - \epsilon) \propto M^{0.12}$. This agrees with the known lack of highly elliptical galaxies for $M \gtrsim 10^{11}\msol$. As $R_e = a\sqrt{1-\epsilon}$, the scaling of mass and roundness explains the tilt in the virial relation that occurs when using $R_e$ instead of $a$ as galaxy scale radius.
\item  Given that (i) $a$ turns out to be the correct proxy for the galactic scale radius and (ii) the best-fit virial relation (Equation~\ref{eq:massplane-a}) fits the data with zero intrinsic scatter, one finds that early-type galaxies are axisymmetric and oblate in general. This agrees with results from modeling their intrinsic shapes based on kinematics and light distributions.
\end{enumerate}



\acknowledgments

I am grateful to Kyu-Hyun Chae (Sejong U) for valuable discussion. This work is based on the \atlas\ database of \citet{cappellari2011, cappellari2013a} and the database of \citet{saglia2016}. I make use of the data analysis software package \textsc{dpuser} developed and maintained by Thomas Ott at MPE Garching  (\url{www.mpe.mpg.de/~ott/dpuser/index.html}). I acknowledge financial support from the National Research Foundation of Korea (NRF) via Basic Research Grant NRF-2015-R1D1A1A-01056807. Last but not least, thanks go to an anonymous referee for helpful comments.


\end{document}